 \documentclass[final,5p,times,twocolumn]{elsarticle}

 \usepackage{graphicx}

\usepackage{amssymb}

 \usepackage{lineno}

 \biboptions{authoryear,semicolon,round,round}

\journal{Science of Total Environment}

\begin{document}

\begin{frontmatter}



\title{A new  FSA approach for in situ $\gamma$-ray spectroscopy}

\title{A new  FSA approach for in situ $\gamma$-ray spectroscopy}

\author[a,b]{A. Caciolli\corref{cor1}}\ead{caciolli@pd.infn.it}
\author[c]{M. Baldoncini}
\author[d]{G. P. Bezzon}
\author[a]{C. Broggini}
\author[d]{G. P. Buso}
\author[b]{I. Callegari}
\author[b]{T. Colonna}
\author[c,d,e]{G. Fiorentini}
\author[b]{E. Guastaldi}
\author[c,e]{F. Mantovani}
\author[b]{G. Massa}
\author[a]{R. Menegazzo}
\author[b]{L. Mou}
\author[a]{C. Rossi Alvarez}
\author[c,e]{M. Shyti}
\author[d]{A. Zanon}
\author[c,e,f]{G. Xhixha}

\address[a]{Istituto Nazionale di Fisica Nucleare (INFN), Padova Section, Via Marzolo 8 - 35131 Padova, Italy. }
 \address[b]{Center for GeoTechnologies, Via Vetri Vecchi, 34 - 52027 San Giovanni Valdarno, Arezzo, Italy.}
 \address[c]{Physics Department, University of Ferrara, Via Saragat, 1 - 44100 Ferrara, Italy.}
 \address[d]{Istituto Nazionale di Fisica Nucleare (INFN), Legnaro National Laboratory, Via dellÕUniversit\`a, 2 - 35020 Legnaro, Padova, Italy.}
\address[e]{Istituto Nazionale di Fisica Nucleare (INFN), Ferrara, Via Saragat, 1 - 44100 Ferrara, Italy.}
\address[f]{Faculty of Forestry Science, Agricultural University of Tirana, Kod\"er Kam\"ez - 1029 Tirana, Albania.}
\cortext[cor1]{Corresponding author. Tel.: +39 049 967 7146.}

\begin{abstract}
An increasing demand of environmental radioactivity monitoring comes both from the scientific community and from the society.
This requires accurate, reliable and fast response preferably from portable radiation detectors. Thanks to recent improvements in the technology,  $\gamma$-spectroscopy with sodium iodide scintillators has been proved to be an excellent tool for in-situ measurements for the identification and quantitative determination of $\gamma$-ray emitting radioisotopes, reducing time and costs.
Both for geological and civil purposes not only $^{40}$K, $^{238}$U, and $^{232}$Th have to be measured, but  there is also a growing interest to determine the abundances of anthropic elements, like $^{137}$Cs and $^{131}$I, which are used to monitor  the effect of nuclear accidents or other human activities. 

The Full Spectrum Analysis (FSA) approach has been chosen to analyze the $\gamma$-spectra. The  Non Negative Least Square  (NNLS)   and the energy calibration adjustment have been implemented in this method for the first time in order to correct the intrinsic problem related with the $\chi ^2$ minimization which could lead to artifacts and non physical results in the analysis.

A new calibration procedure has been developed for the FSA method  by using in situ $\gamma$-spectra instead of calibration pad spectra.
Finally, the new method has been validated  by  acquiring $\gamma$-spectra with a 10.16 cm x 10.16 cm sodium iodide detector  in  80 different sites in the Ombrone basin, in Tuscany. The results from the FSA method have been compared with the laboratory measurements by using HPGe detectors on  soil samples collected in the different sites, showing a satisfactory agreement between them. 
In particular, the $^{137}$Cs  isotopes has been implemented in the analysis since it has been found not negligible during the in-situ measurements.

\end{abstract}


\end{frontmatter}


\section{Introduction}

In situ $\gamma$-ray spectrometry with sodium iodide scintillators is a well developed and consolidated method for radioactive survey (\citealt{Nuccetelli2008,chiozzi, tyler2007}) with a wide range of applications from mineral exploration to environmental radiation monitoring (\citealt{IAEA}), providing quantitative information especially about abundances of principal natural radioisotopes, $^{40}$K, $^{238}$U and $^{232}$Th (\citealt{IAEA1990}).

The experiences of nuclear power plant accidents  and atmospheric nuclear weapon tests taught us that the fallout of man-made radioisotopes ($^{137}$Cs, $^{134}$Cs and $^{131}$I) can affect wide portions of the planet. 
In order to employ such powerful techniques in this context the sensitivity and the quickness have to be improved (\citealt{tyler}).

The conventional approach for studying the specific activity concentration of the three principal natural radioisotopes is to monitor broad spectral windows during the analysis (\citealt{Verdoia,Desbrat}). Generally, these windows are chosen around the photopeaks of $^{40}$K (1460 keV), of $^{214}$Bi (1765 keV), and of $^{208}$Tl (2614 keV). Since the $^{238}$U and the $^{232}$Th are not  $\gamma$-rays emitters their concentrations are evaluated detecting the $\gamma$-rays produced by $^{214}$Bi and $^{208}$Tl respectively. The assumption of secular equilibrium of the decay chains is required in order to use this approach. In addition to the abovementioned radionuclides, the three-windows method has been extended to the measurement of $^{137}$Cs (\citealt{cresswell,Sanderson}). 

The limit of this technique is that it becomes imprecise for short time acquisitions and it suffers the poor intrinsic energetic resolution of NaI(Tl) detector. 
In particular, the Compton continuum around $^{137}$Cs photopeak is generally very intense compared to the intensity of 662 keV $\gamma$-line. This translates into long acquisition times. 
Moreover, the intrinsic energetic resolution of sodium iodide detectors prevents to resolve the triplet at  583 keV ($^{208}$Tl), 609 keV ($^{214}$Bi), and 662 keV ($^{137}$Cs). This issue can be solved only by post processing the data. In any case the windows analysis method requires a prior knowledge of the presence on site of such radioisotope. As a matter of fact, this method is blind to unexpected radionuclides.

Significant improvements in $\gamma$-ray spectrum analysis have been obtained by implementing the full spectrum analysis (FSA) method (\citealt{Hendricks, Minty,Guillot,Gutierrez,tyler2007}). Since the FSA uses the full extent of the spectrum, as opposed to the three windows method, it needs a much lesser statistic to reach the necessary accuracy. This, in turn, means a drastic reduction in acquisition times.

In this paper the non-negative least square (NNLS) constraint and the energy calibration adjustment have been implemented into the FSA algorithm. 
This new approach to the FSA  was applied to a 10.16 cm x 10.16 cm portable NaI(Tl) detector dedicated to in situ $\gamma$-ray spectrometry.

Moreover, we experimented a new approach to the efficiency calibration procedure. Instead of using calibration pads, we used the spectra acquired in sites selected specifically for calibration purposes as described later in this article. 

A detailed characterization of each site (both for calibration and for measurement) was performed by sampling the soil in an area of about ten meters radius and by measuring, in the laboratory, the radioisotopes concentrations by using a high resolution apparatus made of two lead shielded HPGe detectors (called MCA-Rad and described by \citealt{RAD}).  
The NaI(Tl) detector was used within a larger project devoted to the investigation of  the soil characteristics in Ombrone basin (located in Tuscany Region) in morphologic, pedologic, and lithostratigraphic way. 
About 80 different sites have been measured with two main finalities: 
the radioactivity characterization of soil for the study of lithological differences of the altered substratum and the validation of our method.
The results of the new FSA analysis are furthermore  compared to the characterization performed by measurements on sample in laboratory with the MCA-Rad apparatus and to the traditional three windows methods.

\section{Calibration procedure and developments}\label{section2}

Portable instruments are usually calibrated by means of  standard  spectra  acquired  at  least using three  concrete  pads enriched in K, U and Th and a pad free of radioactivity representing the background (\citealt{Hendricks,Engelbrecht,Lovborg}). These pads are usually concrete constructions of cylindrical shape, having finite dimensions of 2-3 m in diameter and 0.3-0.5 m thick and for this reason it is needed a geometrical correction due to the differences from an infinite calibration sources.

Hendriks et al.  proposed as an alternative solution  to build pads by using not concrete but silversand mixed with KCl for the potassium pad, monazite for the thorium pad, and slags for the uranium pad. 

The design of an ideal pad with one radionuclide inside and with a perfect homogeneous distribution of the radioisotope in its volume is impossible, because impurities and non-homogeneities are always present. For example, a clear contamination of uranium in thorium pads has been often seen, as reported by Hendriks and co-workers.
For this reason, compromises between accuracy and applicability of the method have to be weighted. 
In the case of in situ $\gamma$-ray spectrometry accuracies of the order of less than 15\% are usually well accepted which legitimize the above assumptions. 

It is worth  mentioning that the hypothesis of homogeneous distribution of the radionuclides in the  pad mixture should be verified and that the cost of production, handling and storage of the pad is not negligible.

Instead of building pads, an alternative calibration procedure is described in the present  work.
It is based on the selection of sites characterized by a prevalent concentration of one of the natural radionuclides.
Even if it is almost impossible to select one site which contains only one of the nuclides, the selection will be oriented toward sites with unbalanced contents. 
Since  most of the $\gamma$-rays emanating from the earthÕs surface originate in the top 50 cm (\citealt{IAEA}) depending on the rock/soil density the radioactivity characterization of each site is performed sampling at 10 cm depth. 
All calibration sites were selected using geological and geomorphological considerations and further validated trough laboratory measurements (a list of all the sites and the concentration of each isotope are reported in Table \ref{table1}).
The in situ measurements  can be affected by the specificity of the place, like the soil non-homogeneity,  the ground morphology, the non secular equilibrium in radioactive chains, the vertical distribution of $^{137}$Cs, the presence of vegetation, moisture, etc.
Thus, the calibration sites  should be selected according precise prescriptions:
\begin{itemize}
\item relatively uniform distribution of radionuclides in secular equilibrium with their products,
\item plane area without any steps and large enough to be approximated as an infinite source (maximum 10 m radius),
\item undisturbed areas: assuring relatively constant $^{137}$Cs vertical distribution
\item uniform and relatively homogeneous soil moisture content and vegetable coverage.
\end{itemize}
For each site, a variable number of samples from 5 to 12 was  collected in random positions within 10 m  radius in order to check also the homogeneity of the site around the detector since more then 90\% of $\gamma$-rays detected by the sodium iodide are produced by the 7 m radius and 0.5 m deep area around (\citealt{Grasty}). 
The homogeneity of the each site is assured inside the error reported in the in Table \ref{table1}.
Soil and rock samples were dried, homogenized (less than 2 mm fine grain size) and sealed in measurement containers for at least four weeks in order to allow the $^{222}$Rn growth up prior to be measured using high-resolution $\gamma$-ray spectrometry system MCA-Rad following international standards of analysis (\citealt{ASTM,UNI}). 
Each sample is inserted between two HPGe detectors which are shielded with few cm copper and 10 cm lead in order to reduce the environmental radioactivity from the laboratory hall. Nitrogen gas is also fluxed into the lead shielding in order to remove residual radon gas.
The MCA-Rad has been calibrated by using both certified radioactive sources (i.e. $^{152}$Eu, $^{56}$Co) and certified reference materials (RGK-1, RGU-1, and RGTh-1 produced by IAEA) containing known amount of natural radioisotopes of potassium, thorium, and uranium.
These two different calibration procedures are in agreement within their systematics (about 5\%).
Only the CA1 site (discussed in the next section in details) is not a outdoor site, but it is made by a pad of KCl fertilizer. 
The concentrations  reported in Table \ref{table1} are the average of the  measurements performed in the laboratory on the collected rock samples. This way the heterogeneity of each site is properly implemented by the errors  which are dominated by the spread in the results of the  collected samples in each calibration site.
The choice of a number of calibration sites greater than the number of analyzed elements, as in our approach,  is a mandatory to avoid artifacts in the sensitive spectra.
\begin{table}
\centering
\begin{tabular}[htb]{c|c|c|c}
\hline \hline
Site & K [\%] & eU [ppm] & eTh [ppm] \\
\hline
CA1&	53.9$\pm$0.7	&	$<$ 1.0	&	6.0$\pm$0.5$^a$ \\
CC2	& 0.06$\pm$0.02 &	0.7$\pm$0.3	&	0.6$\pm$0.7	\\
GC1	& 0.07$\pm$0.04 &	0.27$\pm$0.08	& 1.14$\pm$0.13	\\
GV1	& 4.9$\pm$0.6	&	7.5$\pm$1.1	&	7$\pm$1	\\
PM2 &	2.26$\pm$0.05 &	2.27$\pm$0.18	& 10.7$\pm$0.8\\
RT1	& 0.10$\pm$0.01	& 6.8	$\pm$1.1	& 	1.74$\pm$0.16	\\
SM1	&1.54$\pm$0.14	& 1.6	$\pm$0.3 &		8.6$\pm$0.6\\
SP2 &	2.92$\pm$0.08 &	7.5$\pm$0.4	&	39$\pm$2	\\
ST2	& 7.8$\pm$0.9	&	36$\pm$5	&	124$\pm$16	\\
\hline
\multicolumn{4}{l}{$^a$ the Th  is obtained by the  procedure explained in section \ref{section3}}\\
\hline \hline
\end{tabular}
\caption{the average of the distribution of natural radioisotopes concentration. The errors correspond to one standard deviation. The conversion factors from Bq/kg are obtained by \citealt{IAEA}: 1\%  = 313 Bq/kg for potassium, 1 ppm = 13.25 Bq/kg for uranium and 1 ppm = 4.06 Bq/kg for thorium.}
\label{table1}
\end{table}

\begin{figure}[!htb]
\begin{center}
\includegraphics[width=\columnwidth]{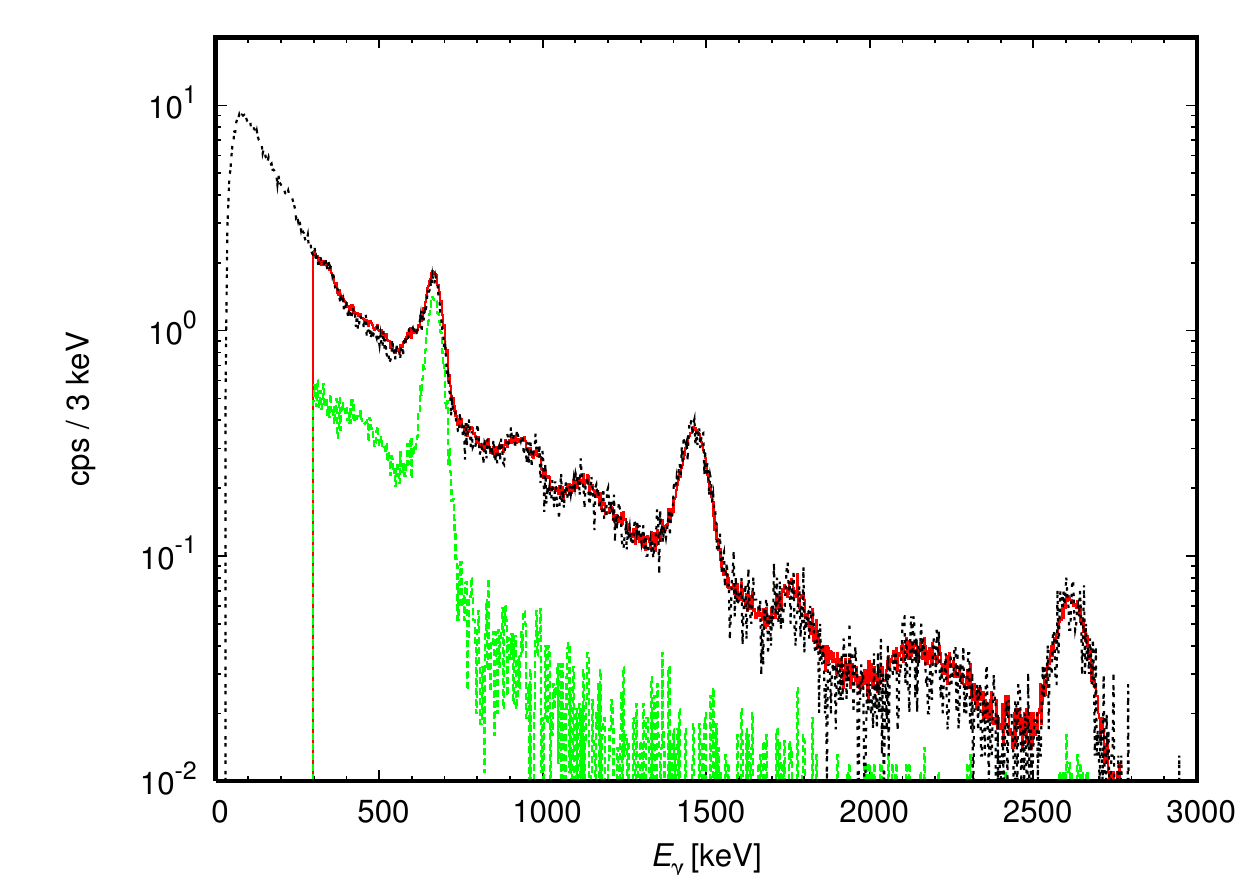}
\end{center}
\caption{spectra acquired with the sodium iodide in situ (black dashed line) compared with the fit obtained by the FSA+NNLS (see section \ref{section3-1}) algorithm (full line red). The $^{137}$Cs contribution is shown alone (green dotted line) to underline the need to include this element in the analysis.}
\label{fig1}
\end{figure}
One advantage of this calibration approach is the possibility to take into account  radionuclides other than $^{40}$K, $^{232}$Th,  and $^{238}$U (in general a minimum of $k+1$ sites are needed, where $k$ is the number of nuclides we want to be sensitive to). 
In our case a site with a prevalent concentration of $^{137}$Cs has been included and used in the calibration (the CC2 site in Table \ref{table1}).
As a matter of fact, after the Chernobyl accident, it is a mandatory to take into account the $^{137}$Cs signal in the spectrum analysis.

This is clearly visible in Figure \ref{fig1} where the presence of cesium  is definitively not negligible (the $\gamma$-peak energy is 662 keV). 
If necessary, other radioisotopes due to nuclear accidents, like $^{131}$I, could be added in the future.

\section{The FSA Algorithm}\label{section3}

The full spectrum analysis method has  been developed in different approaches  (\citealt{H2,Hendricks,Minty,Crossley,Smith})  and was found to be a successful spectrum analysis tool.
When applied to scintillation detectors, it enhances their potentialities for radioisotope quantitative determination. 
As a matter of fact, the FSA  encompasses almost the full energy spectrum, reducing the required statistic of a single measurement and therefore its  duration in time.

The spectra acquired in situ are fitted by a linear combination of the fundamental spectra  derived for each isotope from the calibration analysis. 
The events registered in each channel in the measured spectrum, $N$, can be expressed as:
\begin{equation}\label{eq1}
N(i) = \sum_{k=1}^4 C_k S_k(i) + B(i)
\end{equation}
where 
\begin{itemize}
\item $N(i)$ are the  counts in the channel $i$,
\item $C_k$ are the concentration of the element $k$,
\item $S_k(i)$ are the associated counts to the fundamental spectrum  of the element $k$ in the channel $i$,
\item $B(i)$ are the counts in the channel $i$ due to the intrinsic background. 
\end{itemize}
and the index $k$ stays for $^{40}$K, $^{232}$Th,  $^{238}$U, and $^{137}$Cs. 
It is become a conventional representation for in-situ measurements, for geological purposes,  to express the concentrations of natural radioisotopes in their respective abundances, where $^{40}$K is given in \% weight while eU and eTh are given in ppm. 
The $^{137}$Cs is expressed as the absorbed dose by the detector (nGy/h) due to the heterogeneous  distribution property of anthropic radioisotopes  (described in section \ref{sec:cesium}).
The activity concentrations are deduced applying the least square algorithm to rectangular matrix and  minimizing the $\chi^2$ as in the following equation:
\begin{equation}
\chi^2 = \frac{1}{n-5}\sum_{i=1}^n [N(i) - \sum_{k=1}^4 C_k S_k(i) - B(i)]^2 / N(i),
\end{equation}
where the $N(i)$ is considered poisson distributed and $n$ is   the number of channels in the spectrum used in the analysis (867  in our case).

During the calibration of the system  the fundamental spectra (the $S$ matrix) are obtained by solving equation \ref{eq1} with  the radionuclide concentrations (the $C_k$ coefficients) reported in Table \ref{table1}:
\begin{equation}
[S] = [C]^{-1} \times [N].
\end{equation}
It has to be noted that FSA calibration method produces also the intrinsic background, $B$,  which can be compared with a spectrum acquired with the detector inside a thick lead shielding.

Only the energy range from 300 keV up to 2900 keV is considered in the analysis.
Below 300 keV  there is a strong presence of the backscattering events which depends  on the atomic number and density of the surrounding materials. 
Above 2900 keV only the cosmic ray contribution is present\footnote{The cosmic background has been found to be almost constant in our measurements and negligible with respect to the intrinsic background of the NaI(Tl) detector. However, its contribution is averaged and taken into account by the $B$ sensitive spectrum determined by the algorithm.}.

Once the first solution has been obtained, in order to improve the $\chi ^2$ minimization, a trimming procedure is executed by changing the site concentrations ($C_k$) in small steps around the measured intervals and repeating the matrix solution.
\begin{figure}[h]
\begin{center}
\includegraphics[width=\columnwidth]{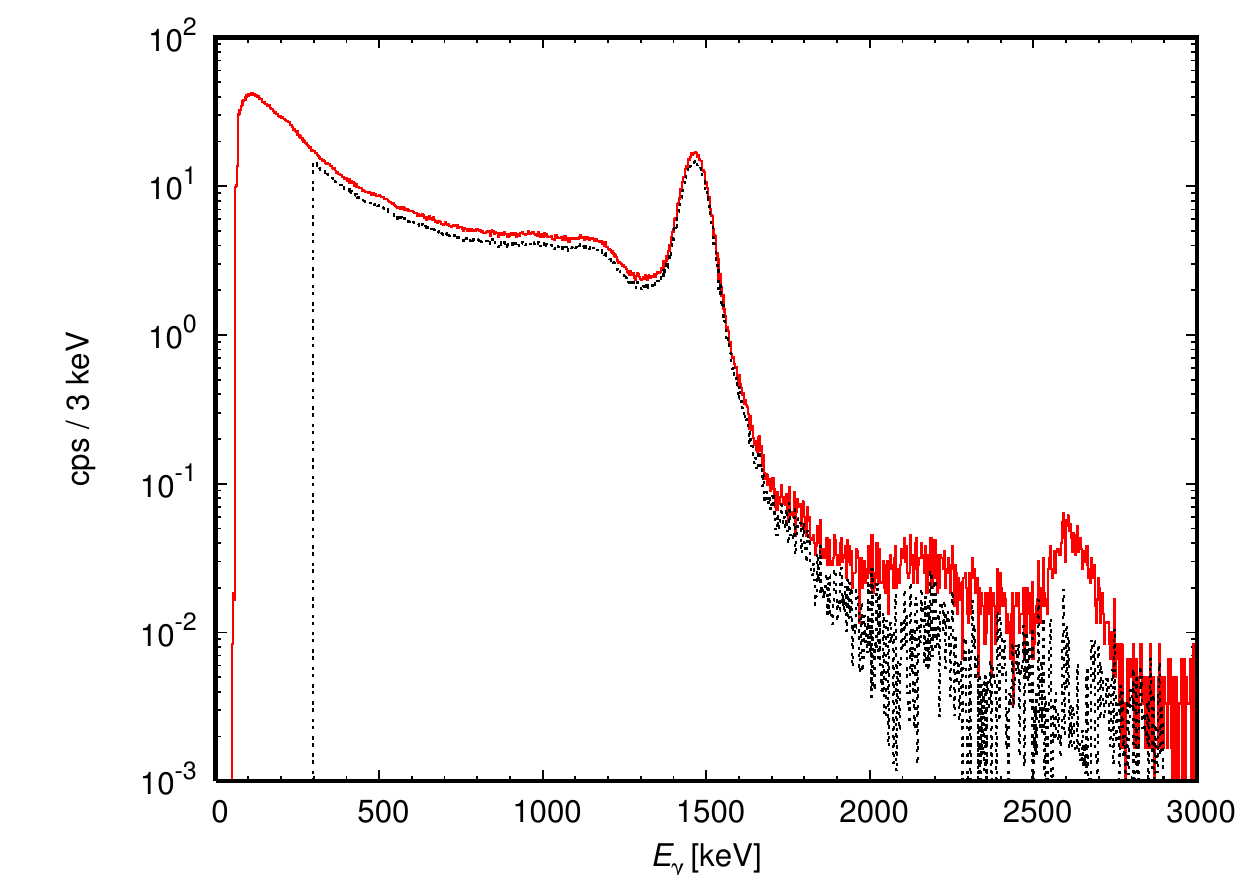}
\end{center}
\caption{the spectra acquired in CA1 site in (red full line). The fit obtained using the concentrations measured with the MCA-Rad system is also reported (green dashed line).}
\label{fig2}
\end{figure}
This strategy has been implemented in order to correct the unavoidable differences between the sample measured concentrations and the average site values.
The need of this correction is  evident observing the spectra acquired at the CA1 site where the sodium iodide detector was placed on a large platform made of KCl fertilizer used as potassium calibration site. 
The KCl was stored inside a building and in the NaI(Tl) spectra are clearly visible the peaks due to the thorium and uranium contained inside the building walls.
In the samples analyzed by the MCA-Rad system, the contribution from these two elements was not present. 
Figure \ref{fig2} shows the $\gamma$-spectrum acquired in CA1 site compared with the fit spectrum obtained by assuming  only the concentrations obtained by  the measurements on samples in laboratory.
Except for CA1, this procedure applied to the other sites produce a deviation from the concentrations reported in Table \ref{table1} within 1$\sigma$ which confirms the assumption of acceptable homogeneity declared in the previous section.

\subsection{Improvements in FSA}\label{section3-1}

\begin{figure}[h]
\begin{center}
\includegraphics[width=\columnwidth]{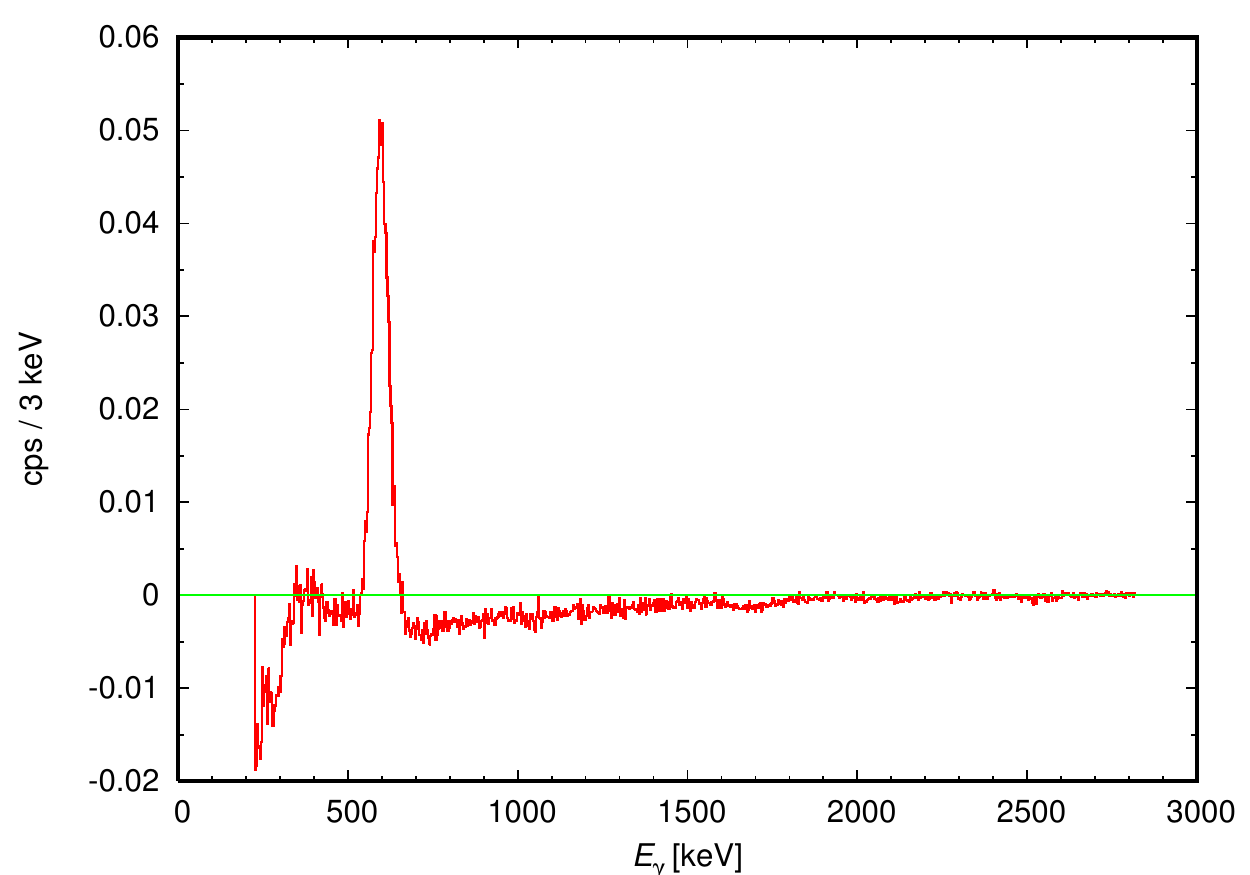}
\end{center}
\caption{the sensitive spectra of $^{137}$Cs, obtained using the standard FSA. The green line is placed to show the zero counts level.}
\label{fig3}
\end{figure}

\begin{figure}[h]
\begin{center}
\includegraphics[width=\columnwidth]{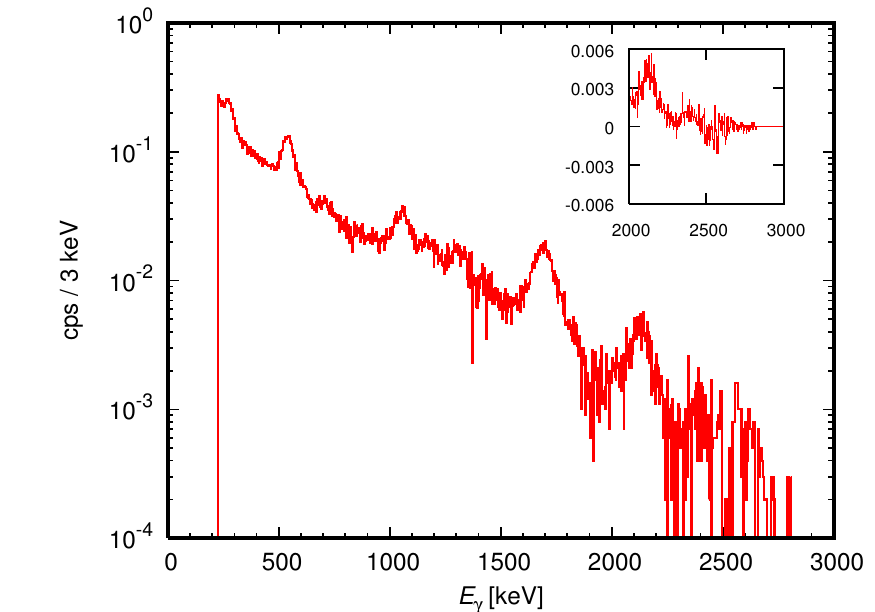}
\end{center}
\caption{the sensitive spectra of $^{238}$U, obtained using the standard FSA. The region where there are negative counts is emphasized in the box.}
\label{fig4}
\end{figure}

The $\chi ^2$ minimization without any further conditions, which is the base of the FSA method, can bring  to sensitive spectra having energy regions of negative counts.
Two evident examples of this problem are shown in Figure \ref{fig3} and \ref{fig4}.

The presence of these non physical results introduces crosstalk effects in the analysis, leading to systematic errors.
The NNLS (Non Negative Least Square) constraint (\citealt{NNLS0,NNLS1,NNLS2}), which forces the counts on each bin to be zero or positive, has been for the first time implemented in the FSA algorithm in order to avoid this problem (\citealt{marica}).
The NNLS algorithm is based on the Kuhn-Tucker condition for Problem of Least Square with Linear Inequality Constraint (LSI) (\citealt{NNLS0}). In the case a problem has a solution with least square minimization this theorem states that there is also a solution of the problem with non negative solution.

\begin{figure*}[htb]
\begin{center}
\includegraphics[width=\textwidth]{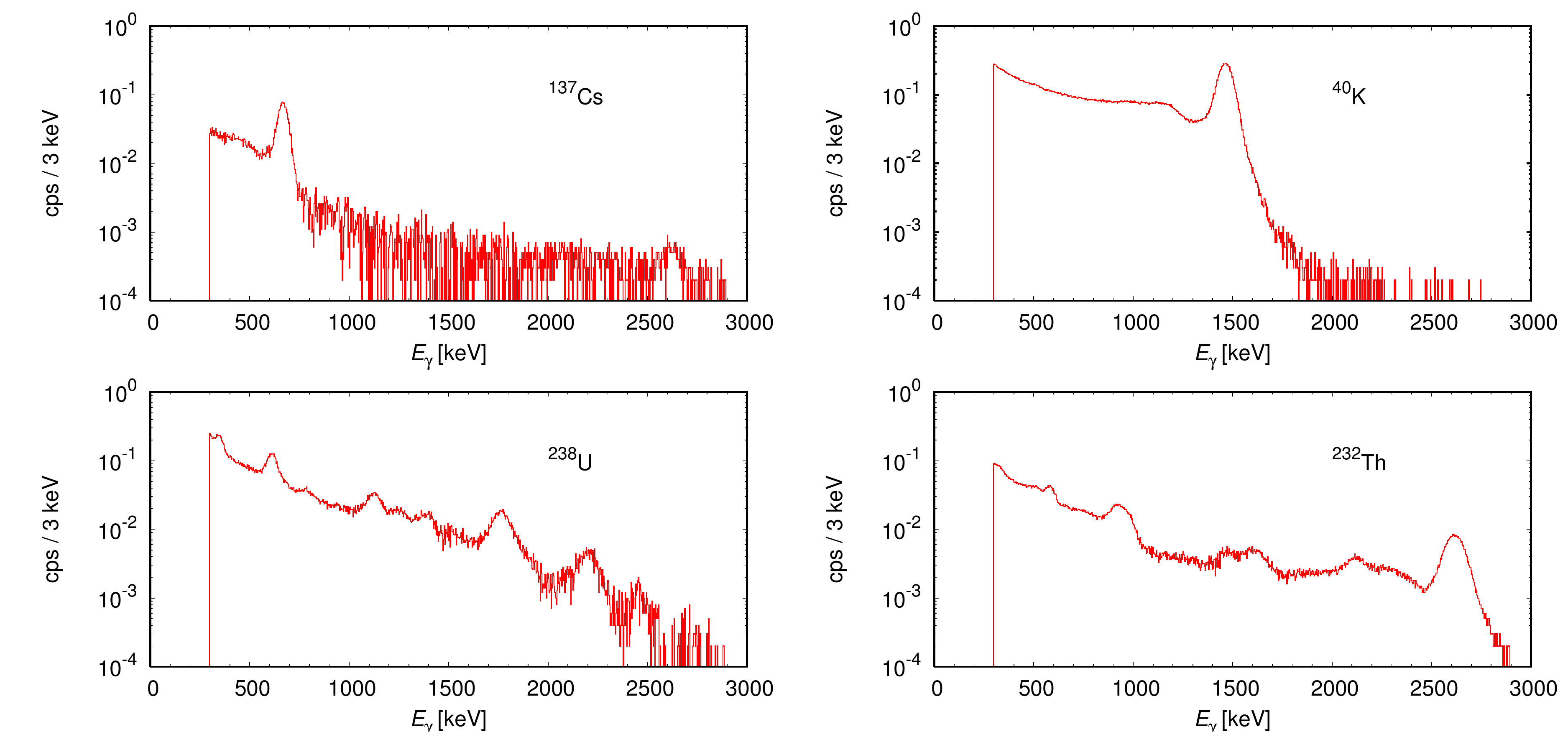}
\end{center}
\caption{the sensitive spectra obtained through the FSA with NNLS constraint.}
\label{fig5}
\end{figure*}

The sensitive spectra calculated with the new algorithm are shown in Figure \ref{fig5}, where it can be directly seen a more reliable sensitive spectra with the NNLS implementation.

It has to be noted that the sensitive spectra resolution can be severely affected by gain mismatch between the different site spectra, which has to be always calibrate properly. 
Finally, residual correlations between isotopes are still present in the sensitive spectra shape, due to the presence of all the  radionuclides in most of the selected sites. 

\subsection{$^{137}$Cs}\label{sec:cesium}

The $^{137}$Cs has been added to the analysis since it was found in not negligible quantities in soils. It was produced by the Chernobyl accident and randomly deposited by climatic events in Italy regions.
The profile distribution of $^{137}$Cs in soil systems shows a tendency of decrease toward deeper layers (\citealt{Z}).
The $^{137}$Cs is also heterogeneous in soils and this is confirmed by the measurements performed on collected samples which show a large distribution of $^{137}$Cs values on the surface (see Table \ref{cesiumtab}).
As a matter of fact, the NaI(Tl) detector, which records the $\gamma$-rays produced by a wide surface area, is the best solution in order to average the cesium amount in the soil. In particular, it is possible to  avoid over/under estimations due to point-like sampling.
It has to be noted that the sensitive spectrum for $^{137}$Cs is dominated by the CC2 site which is poor of natural radioisotopes and with an high value of $^{137}$Cs.
This way the cesium sensitive spectrum is free from correlations with the other radioisotopes.
The dose absorbed by the detector, $D$, which takes into account  the heterogeneity of the isotope in soils, is calculated summing the counts in each channel of the sensitive spectrum $S_{Cs}(i)$ weighted with the channel's energy $E(i)$ and detector mass, $m$. This is multiplied by the $C_{Cs}$ coefficient derived by the FSA+NNLS algorithm as in the following equation:
\begin{equation}
D = \frac{1}{m}C_{Cs}\sum_i^n S_{Cs}(i)E(i),
\end{equation}
As described in Section \ref{section2}, the sensitive spectra are calculated in a range of energy from 300 kev up to 2900 keV. The contribution to the dose due to energies below 300 keV has been included a posteriori by measuring the signal of a $^{137}$Cs radioactive source and it brings a 5\% systematic uncertainty.
\begin{table}
\centering
\begin{tabular}[!htb]{cccc}
\hline \hline
site &  Bq/kg & site & Bq/kg	\\
\hline									
CC2	&	8	$\pm$	4	&GC1	&	0.87	$\pm$	0.04 \\ 
 RT1	&	6	$\pm$	5	& SM1	&	$26^{+37}_{-26}$	\\
ST2 & $61^{+100}_{-61}$ & SP2	&	23	$\pm$	2	  \\
GV1 & 31$\pm$18 & PM2	&	18	$\pm$	9 \\
\hline \hline
\end{tabular}
\caption{$^{137}$Cs concentrations recorded on the collected samples in each calibration site. The results of the samples are averaged and the errors are 1$\sigma$ standard deviation.}
\label{cesiumtab}
\end{table}

\section{Spectra Analysis and Method Validation} \label{section4}

\begin{figure}[!htb]
\begin{center}
\includegraphics[width=\columnwidth]{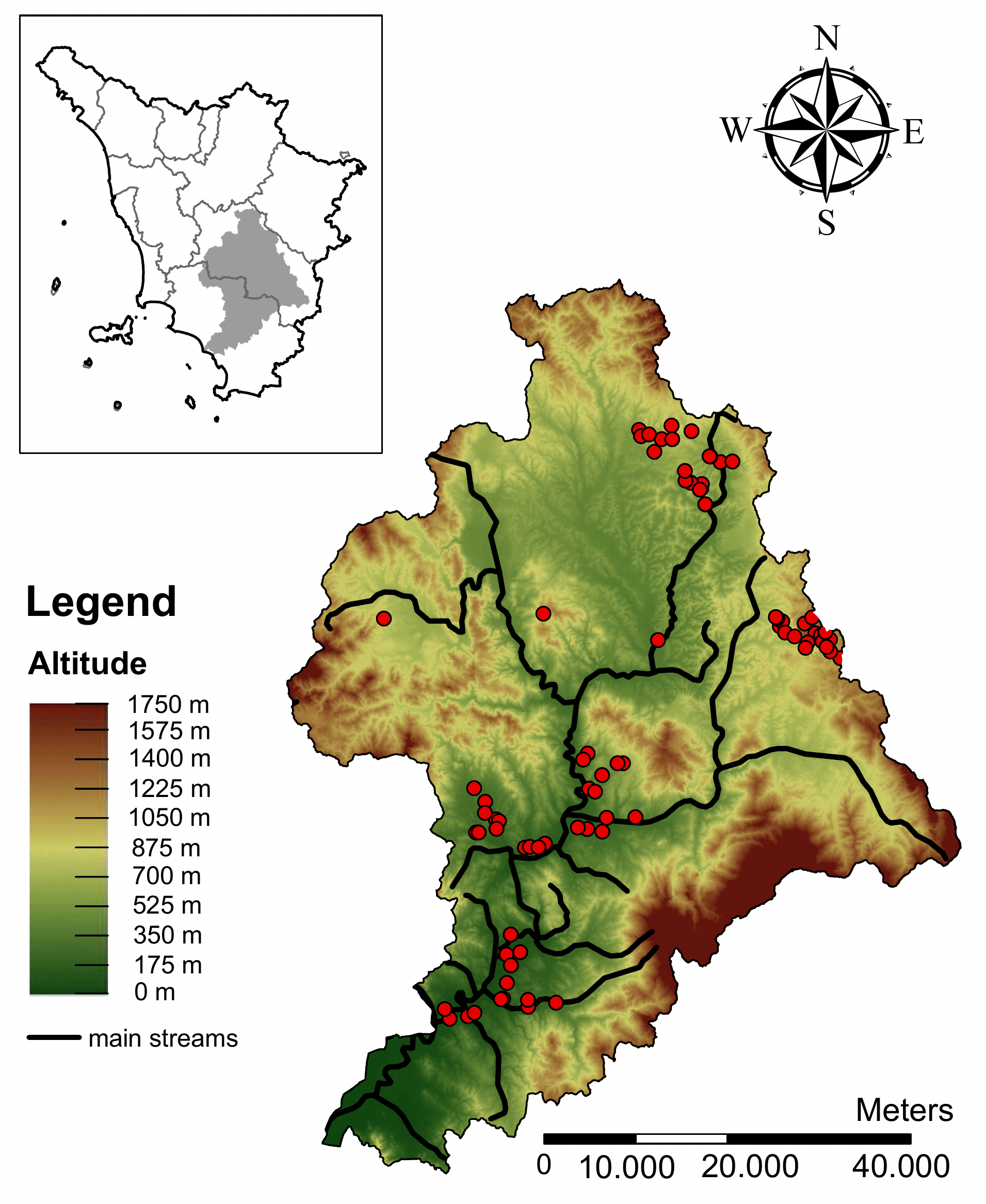}
\end{center}\caption{the position of the 80 sites used for the validation (red circles) in the area of Ombrone basin.}
\label{ombrone}
\end{figure}

Recently an intense campaign of measurements was dedicated to the geological study of the soils in Tuscany, in an area investigated is situated between the regions of Siena and Grosseto in the Ombrone's basin (see Figure \ref{ombrone}). 
This territory is characterized by a highly diversified geological structure and it has been studied by measuring different aspects and characteristics of the soil.
The natural radioactivity of this area has been also measured over  80 sites, where it has been investigated both by acquiring spectra in situ with the NaI(Tl) and by collecting samples for measurements with  the MCA-Rad system.
The sites selection fulfilled the same characteristics chosen for the calibration procedure, in particular for the topology of the site.
For each site the 10.16 cm x 10.16 cm NaI(Tl) detector was used to perform 5 minutes ground measurements in-situ: the acquisition time was chosen in order to have enough statistics for a comparison with the three widow method too.
 Each spectra was analyzed using FSA method with and without NNLS constrain (based on the new calibration approach described in section \ref{section2}) together with three windows method. In each site were collected one sample in the position where was grounded the detector and four samples were collected on the side bisectors of a 2 m side square centered at the grounded position of the detector. Each sample is treated in the same way as the one collected for the efficiency calibration of the system. The averages of the results obtained form measurements in laboratory for each site was used as reference to be compared with the outputs of the two methods. 
\begin{table}
\centering
\begin{tabular}[!htb]{cccc|c}
\hline \hline
 & K 	& U	&	Th	& $\chi ^2$ \\
\hline									
IAEA &1.12$\pm$0.07&1.11$\pm$0.10 &	1.00$\pm$0.09 & ---	\\	
FSA	&0.99$\pm$0.06& 0.78$\pm$0.14 &	0.86$\pm$0.07	 & 1.22$\pm$0.08 \\
NNLS	&1.00$\pm$0.06&	0.82$\pm$0.13&	0.92$\pm$0.07&	1.08$\pm$0.08 \\
NNLS opt.	 &1.06$\pm$0.06&	0.87$\pm$0.12&	0.94$\pm$0.07&	1.06$\pm$0.05\\
\hline \hline
\end{tabular}
\caption{the $\Omega$ coefficients averaged for all the data samples. For the IAEA results the $\chi ^2$ is not shown due to the absence of a fit procedure.}
\label{table2}
\end{table}

In Table \ref{table2} the correlation factor, $\Omega$, which minimize the relative dispersion is obtained  by using the following  equation:
\begin{equation}\label{eq2}
y = \sum_{i=1}^{80} \frac{(NaI_i - \Omega MCA_i)^2}{MCA_i^2}
\end{equation}
where, $MCA$ and $NaI$ are referred to the radioisotopes concentrations calculated in laboratory and in situ. 
There is an agreement within the uncertainties between the FSA and the window method, but it seems that the window method usually  overestimates the concentration while both FSA versions go in the opposite direction.
For the FSA method three different algorithms  have been compared and the results are reported in Table \ref{table2}: the standard FSA described in literature, the data obtained implementing the NNLS in the algorithm, and the FSA+NNLS with the optimization procedure introduced in the calibration method.
This way, it is possible to understand the effect on the resulting isotope concentrations for each step by using the $\chi ^2$ as reference.
All results reported in Table \ref{table2} agree with a factor $\Omega = 1$, which guaranties the reliability of the method  for all elements. The correlation for the uranium element is affected by the  atmospheric radon concentration at the time of the in situ measurement, although this discrepancy is within the uncertainties.

\begin{figure}[!htb]
\begin{center}
\includegraphics[width=\columnwidth]{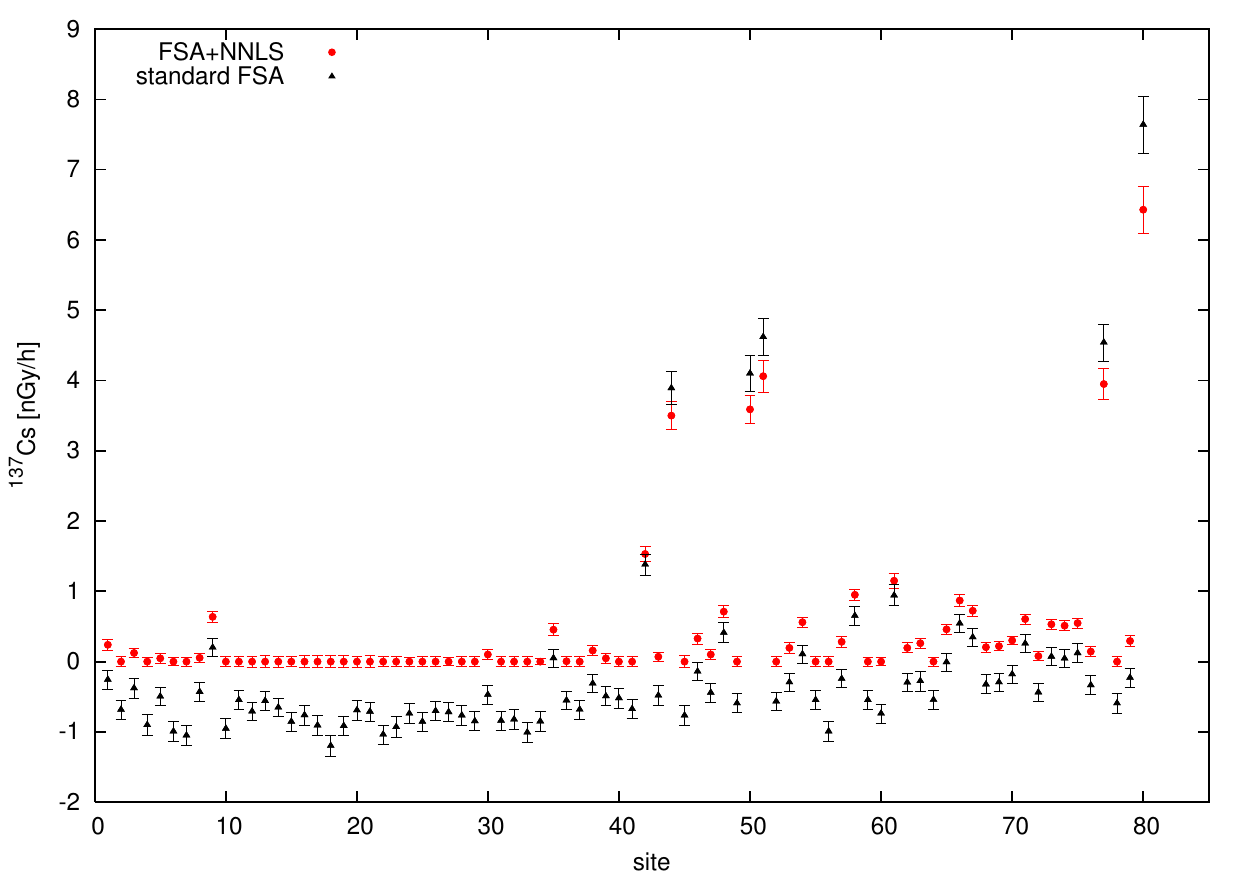}
\end{center}\caption{the measured $^{137}$Cs dose in the 80 sites determined with the new FSA+NNLS approach (red circles) and with the standard FSA (black triangles). The new algorithm avoids the negative counting and it reduces at the same time the uncertainties.}
\label{fig6}
\end{figure}

The energy calibration adjustment has been included in the analysis of the measured spectra in order to correct for possible gain mismatching between the measured spectrum and the sensitive one.
This problem is minimized implementing this post process correction as proven  by the reduction of the  $\chi ^2$ reported in Table \ref{table2}.
As an example, in Figure \ref{fig6} the measured $^{137}$Cs concentrations, determined using  both  the standard FSA and in the FSA+NNLS methods, are reported. When the cesium abundances are low, the standard FSA algorithm introduces negative concentrations in order to have a better fit  of the spectrum shape (corresponding to the $^{137}$Cs line there is also the 609 keV peak due to the $^{214}$Bi). 

Introducing the NNLS and the gain drift,  the quality of fits is improved by 10\% in average as shown in column 5 in Table \ref{table2}.
\begin{figure}[!htb]
\begin{center}
\includegraphics[width=\columnwidth]{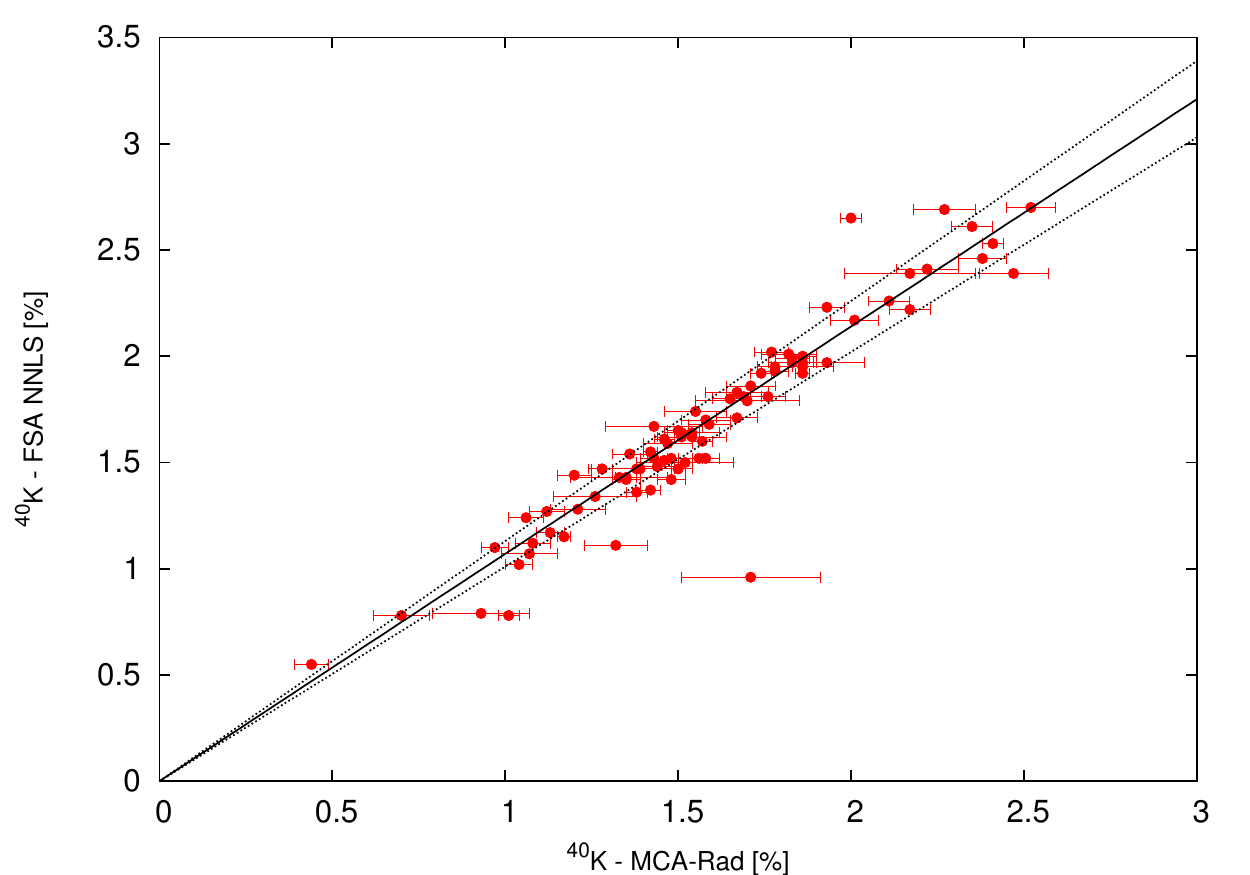}
\end{center}\caption{potassium concentration measurements done with the MCA-Rad system and with the FSA+NNLS analysis are plotted together. The average of five soil samples with relative uncertainty is plotted for the MCA-Rad analysis. No errors are associated with the NaI(Tl) data. The linear correlation line associated with the $\Omega$ value is shown (full line) with 1 $\sigma$ error (dashed line).}
\label{fig7}
\end{figure}
\begin{figure}[!htb]
\begin{center}
\includegraphics[width=\columnwidth]{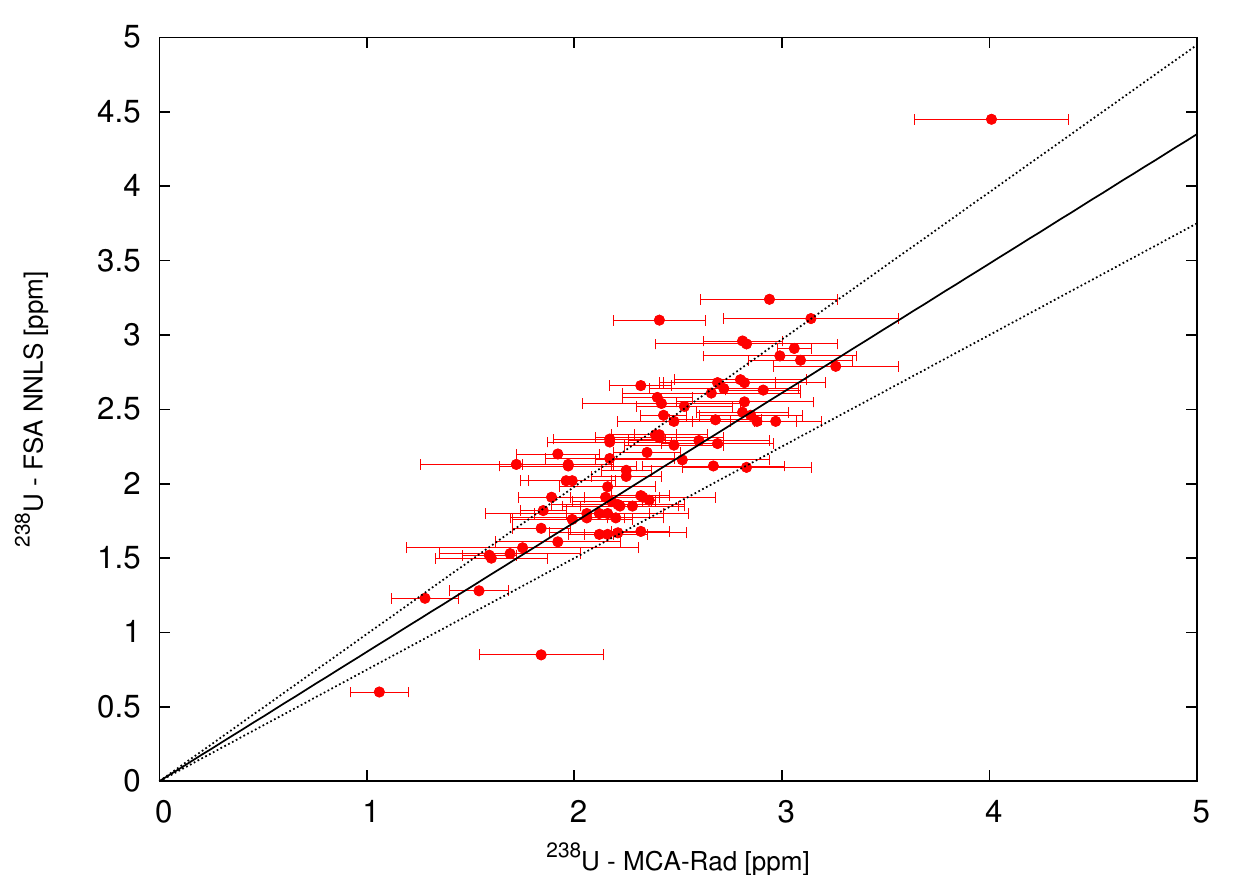}
\end{center}\caption{uranium concentration measurements done with the MCA-Rad system and with the FSA+NNLS analysis are plotted together. The average of five soil samples with relative uncertainty is plotted for the MCA-Rad analysis. No errors are associated with the NaI(Tl) data. The linear correlation line associated with the $\Omega$ value is shown (full line) with 1 $\sigma$ error (dashed line).}
\label{fig8}
\end{figure}
\begin{figure}[!htb]
\begin{center}
\includegraphics[width=\columnwidth]{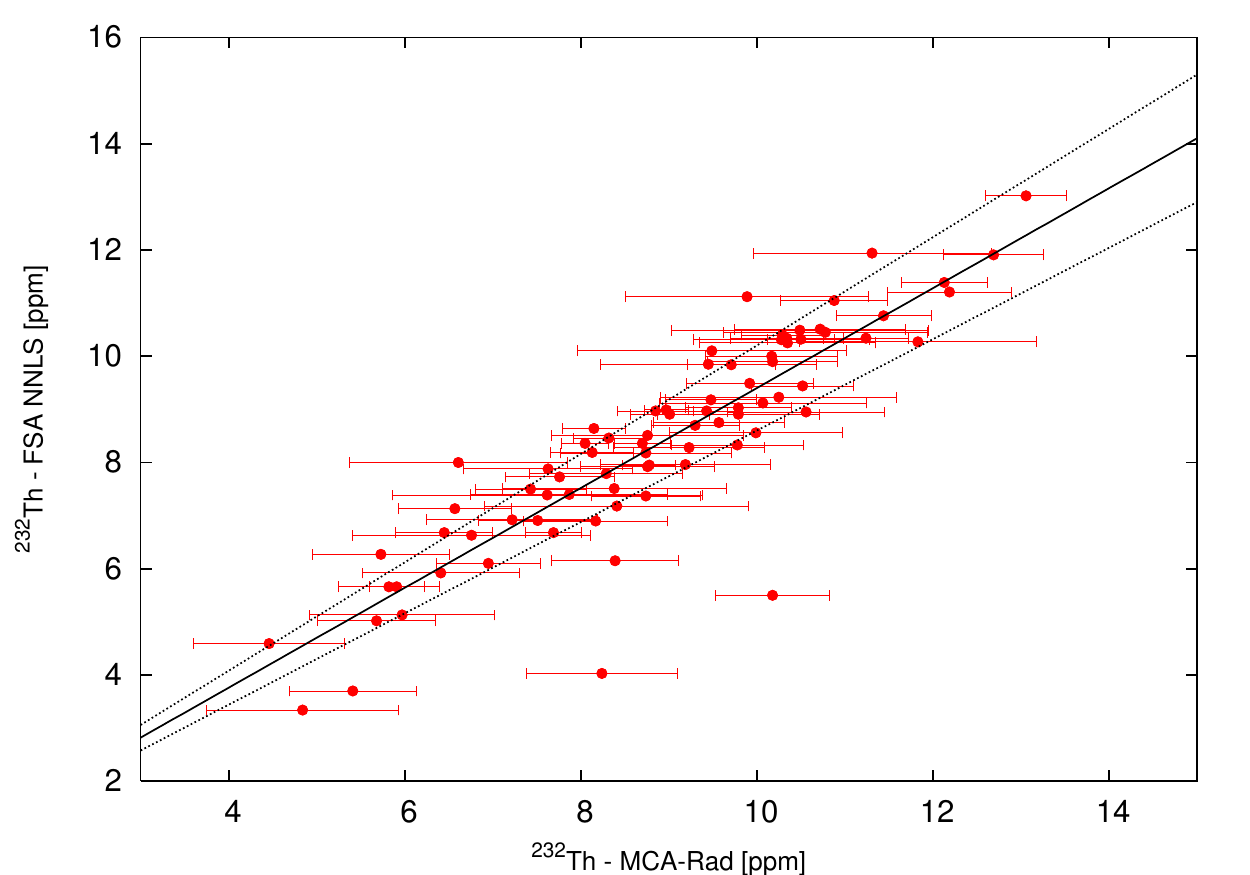}
\end{center}
\caption{thorium concentration measurements done with the MCA-Rad system and with the FSA+NNLS analysis are plotted together. The average of five soil samples with relative uncertainty is plotted for the MCA-Rad analysis. No errors are associated with the NaI(Tl) data. The linear correlation line associated with the $\Omega$ value is shown (full line) with 1 $\sigma$ error (dashed line).}
\label{fig9}
\end{figure}

The correlation between the in situ measurements, analyzed by the FSA+NNLS, and the MCA-Rad  measurements for all 80 sites are reported in  Figure \ref{fig7}, \ref{fig8}, and \ref{fig9}. The uncertainty reported on the MCA-Rad is the standard deviation of  the average calculated over the five collected samples.

The uncertainties on the $\Omega$ factors are used as systematic uncertainties on the concentration measurements with the sodium iodide, since they contain both the contribution from the non homogeneity of soil ground and the systematics due to the analysis algorithm.
The  uncertainties found in this way are: 5\% for the potassium, 14\% for the uranium and 7\% for thorium, which are smaller than the ones requested for outdoor in situ studies, for civil and also geological purposes.

\section{Summary and Conclusions}
 
 The in situ $\gamma$-ray spectroscopy with sodium iodide scintillators is one of the most powerful techniques for the measurement of radionuclides concentration.
 The main advantages are its modest time consumption, its portability, and the reasonable cost of the detector with respect to competing systems (\citealt{Saidou}).
 These characteristics make the use of sodium iodide scintillator a prime candidate to meet the recent increase in the demand for in situ $\gamma$-ray spectroscopy.
  
The FSA is a powerful tool for $\gamma$-spectra analysis thanks to its reduction in required statistic and its increase in analyzable radionuclides.  
It is now also applied to airborne and in situ $\gamma$-ray spectrometry.

Nevertheless, the FSA shows limits due to the intrinsic $\chi ^2$ minimization which brings to non physical results, both in the fundamental spectra construction and in the determination of the element concentrations. 
These problems have been solved in the present work  by introducing the Non Negative Least Square constraint into the FSA algorithm. More reliable fundamental spectra have been obtained and the uncertainties in the fit procedure have been reduced.

At the same time,  a new calibration procedure has been investigated to reduce the difficulties  connected with pads calibration.
This way,  $^{137}$Cs isotope has been also introduced in  the analysis taking into account the anthropic  effect on the environment which is clearly not negligible.

The new algorithm has been validated with measurements performed in 80 sites in Tuscany by using a 10.16 cm x 10.16 cm  NaI(Tl) detector.
In particular the concentrations have been obtained with  a 5\% error on $^{40}$K, 7\% error on $^{232}$Th, and 15\% error  on $^{238}$U concentration.

 The FSA+NNLS method has been used also to analyze airborne measurements where the need of advantage of working with reduced statistics allows to reduced equipment weight and flying costs.
 This analysis approach has been also found to be the optimal solution in order to characterize large amounts of different sites avoiding the necessary of sampling measurements reducing the analysis time and avoiding the treatment of samples.
These results will be the subject of future publications.

\section*{Acknowledgements}
The authors would like to thank Enrico Bellotti, Di Carlo Giuseppe, Pirro Altair, Luigi Carmignani, and Riccardo Vannucci for useful suggestions and invaluable discussions.  This work was partially supported by INFN (Italy) and by Fondazione Cassa di Risparmio di Padova e Rovigo

\bibliographystyle{agsm}
\rule{5cm}{0.01cm}












\end{document}